\documentclass[aps,pra,shownopacs,twocolumn]{revtex4}

\usepackage{siunitx}
\usepackage{ifpdf}
\usepackage{float}
\usepackage{upgreek}
\usepackage{xcolor}
\usepackage{graphicx}
\usepackage{hyperref}
\usepackage{amsmath}
\usepackage{epstopdf}
\usepackage{bbm}
\usepackage{bbold}

\newcommand{\bra}[1]{\langle #1 |}
\newcommand{\ket}[1]{| #1 \rangle}

\usepackage{dsfont}

\begin{document}

\title{A quantum-bit encoding converter}

\author{Tom Darras}
\affiliation{Laboratoire Kastler Brossel, Sorbonne Universit\'e, CNRS, ENS-Universit\'e PSL, Coll\`ege de France, 4 Place
Jussieu, 75005 Paris, France}
\author{Beate Elisabeth Asenbeck}
\affiliation{Laboratoire Kastler Brossel, Sorbonne Universit\'e, CNRS, ENS-Universit\'e PSL, Coll\`ege de France, 4 Place
Jussieu, 75005 Paris, France}
\author{Giovanni Guccione\footnotemark[2]\footnotetext{\footnotemark[2]Present address: Centre for Quantum Computation and Communication Technology, Department of Quantum Science and Technology, Research School of Physics, The Australian National University, Canberra ACT 2601, Australia.}}
\affiliation{Laboratoire Kastler Brossel, Sorbonne Universit\'e, CNRS, ENS-Universit\'e PSL, Coll\`ege de France, 4 Place
Jussieu, 75005 Paris, France}
\author{Adrien Cavaill\`{e}s\footnotemark[3]\footnotetext{\footnotemark[3]Present address: LightOn, 5 Impasse Reille, 75014 Paris, France.}
\affiliation{Laboratoire Kastler Brossel, Sorbonne Universit\'e, CNRS, ENS-Universit\'e PSL, Coll\`ege de France, 4 Place
Jussieu, 75005 Paris, France }}
\author{Hanna Le Jeannic}
\affiliation{Laboratoire Kastler Brossel, Sorbonne Universit\'e, CNRS, ENS-Universit\'e PSL, Coll\`ege de France, 4 Place
Jussieu, 75005 Paris, France}
\author{Julien Laurat}
\email{julien.laurat@sorbonne-universite.fr}
\affiliation{Laboratoire Kastler Brossel, Sorbonne Universit\'e, CNRS, ENS-Universit\'e PSL, Coll\`ege de France, 4 Place
Jussieu, 75005 Paris, France}


\maketitle

\textbf{From telecommunication to computing architectures, the realm of classical information hinges on converter technology to enable the exchange of data between digital and analog formats, a process now routinely performed across a variety of electronic devices. A similar exigency exists as well in quantum information technology where different frameworks are being developed for quantum computing, communication, and sensing \cite{Kimble, Wehner:2018:Science}. Thus, efficient quantum interconnects are a major need to bring these parallel approaches together and scale up quantum information systems \cite{PRXroadmap_interconnect}. So far, however, the conversion between different optical quantum-bit encodings has remained challenging due to the difficulty of preserving fragile quantum superpositions and the demanding requirements for postselection-free implementations. Here we demonstrate such a conversion of quantum information between the two main paradigms, namely discrete- and continuous-variable qubits \cite{Andersen:2015:NatPhys,Minzioni:2019:JOpt}. We certify the protocol on a complete set of single-photon qubits, successfully converting them to cat-state qubits with fidelities exceeding the classical limit. Our result demonstrates an essential tool for enabling interconnected quantum devices and architectures with enhanced versatility and scalability.}

To advance quantum information science and technologies, many quantum machines and protocols are being developed based on an assortment of physical systems, including photons, atoms, mechanical oscillators, solid-state or superconducting devices \cite{Kurizki2015,Pirandola2016}. These implementations draw on different forms of quantum information encoding, depending on the favoured degrees of freedom and the advantages they can provide. For instance, in superconducting quantum computing, discrete-level computational bases and Schr\"odinger cat state encodings are both actively pursued \cite{reviewsuperconducting}. Similarly in quantum photonics, discrete- and continuous-variable approaches are also being successfully leveraged in parallel \cite{Braunstein:2005:RevModPhys,OBrien}.  

With the diversity of platforms, the capability to convert between different encodings is key to linking separate devices in future quantum information processing infrastructures \cite{PRXroadmap_interconnect}. Such a process is challenging as it requires high efficiencies to preserve fragile quantum superpositions and maintain amplitude and phase during the transcoding. Moreover, a quintessential provision is for it to function as a black box with propagating input and output qubits, a requisite that demands the process to be heralded and free of post-selection. Photons are best suited to transmit information and strong candidates to realize such functionality via teleportation \cite{teleport}, locally or at large distance. Recent research in hybrid photonics, where discrete- and continuous-variable approaches have started to be combined for further capabilities \cite{Andersen:2015:NatPhys,Pirandola2016,Minzioni:2019:JOpt,Lee2012,arxivLvovsky}, opened the path to this realization.

Specifically, quantum state engineering experiments have first led to the generation of hybrid entanglement between particle- and wave-like optical qubits \cite{Morin:2014:NatPhot, Jeong:2014:NatPhot,Huang:2019:NewJPhys}, a key resource for such state transfer. This CV-DV entanglement has enabled the remote state preparation of qubit states \cite{Ulanov:2017:PhysRevLett,Le-Jeannic:2018:Optica} and the violation of steering inequalities for cryptography applications \cite{Cavailles:2018:PhysRevLett}. Entanglement swapping was recently demonstrated \cite{guccione2020connecting}, providing the capability to distribute such entanglement between distant hetereogenous quantum nodes. A first teleportation experiment was also implemented \cite{Sychev:2018:NatComm}, albeit with a postselected scheme that prevents its use in any practical application. The realization of a qubit converter has so far proven difficult owing to additional hurdles associated with the independent generation of all the required high-purity resources and the subsequent operations for the conversion process. 

\begin{figure}[t!]
\vspace{-1.3cm}
\includegraphics[width=0.97\columnwidth]{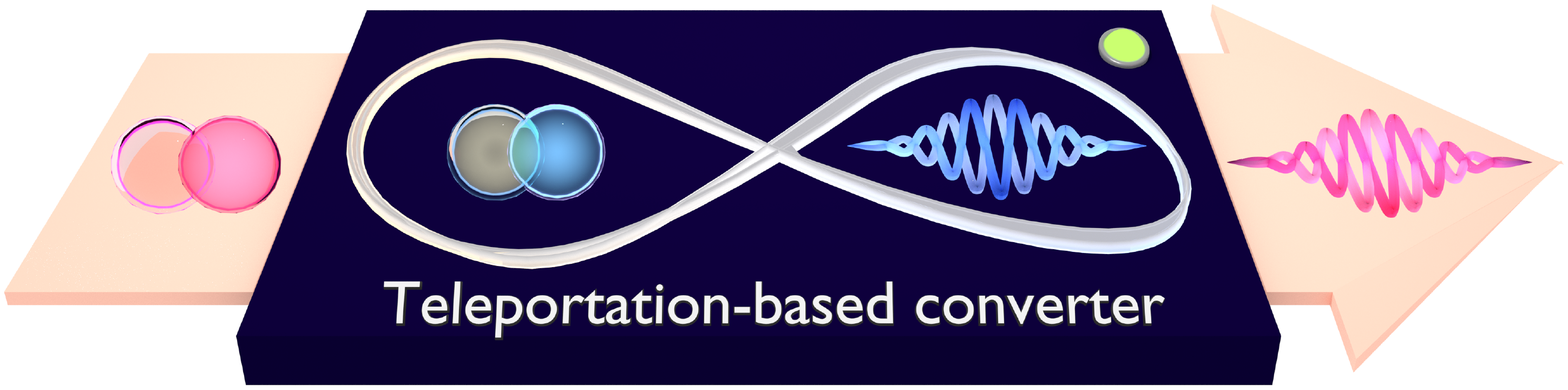}
\vspace{-1.5cm}
\caption{\textbf{An optical encoding converter for quantum interconnects.} A discrete-variable qubit encoded in the Fock-state basis $\ket{\psi}_{DV}=c_{0}\ket{0}+c_{1}e^{i\theta}\ket{1}$ is converted to a continuous-variable qubit encoded in the cat-state basis $\ket{\psi}_{CV}=c_{0}\ket{cat+}+c_{1}e^{i\theta}\ket{cat-}$. The heralded and postselection-free process is based on quantum teleportation via optical hybrid entanglement. The converter enables the exchange of the encoding basis while preserving the superposition, thereby allowing quantum data transfer between diverse systems and platforms.}
\label{fig1}
\end{figure}

Here we overcome these limitations and demonstrate the conversion of optical DV qubits to CV qubits. Our protocol is illustrated in Fig. \ref{fig1}. A DV qubit encoded in the photon-number basis $\{\ket{0},\ket{1}\}$ is converted to a CV qubit in the cat-state basis $\{\ket{cat+},\ket{cat-}\}$, where $\ket{cat\pm}\propto \ket{\alpha}\pm\ket{-\alpha}$ with $\ket{\alpha}$ a coherent state of amplitude $\alpha$. This process is performed by teleportation using optical hybrid entanglement. Importantly, our heralded protocol is postselection free, and the output CV qubit is freely emerging and available for further operations. This apparatus constitutes thereby a first bona fide encoding converter between qubits of different nature. 

\begin{figure*}[t!]
\includegraphics[width=1.89\columnwidth]{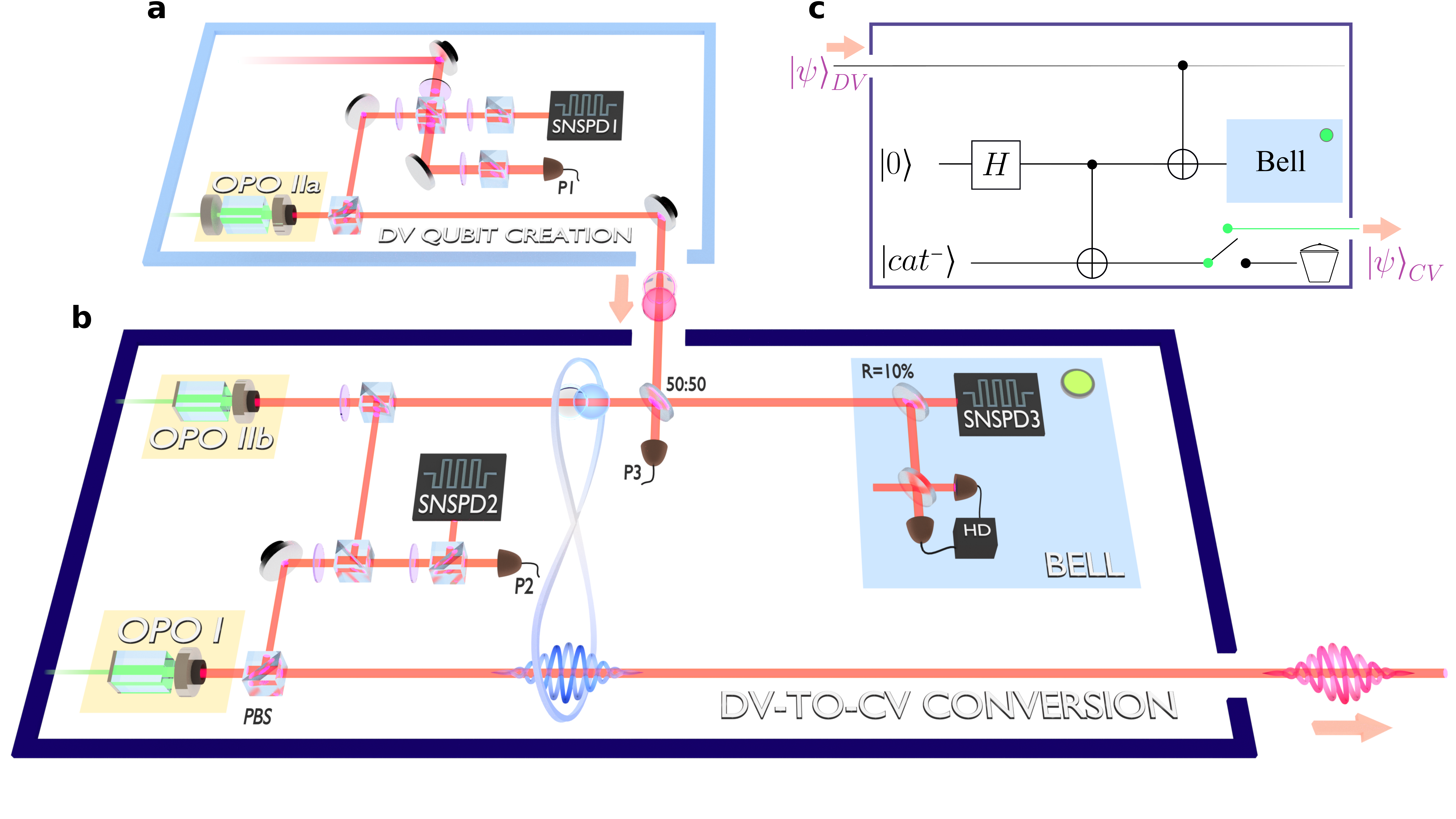}
\caption{\textbf{Experimental setup.} \textbf{a,} A discrete-variable qubit is first created from a two-mode squeezed state generated by an optical parametric oscillator (OPO IIa). The preparation consists in applying a tunable phase-sensitive displacement on the idler mode and subsequent heralding by a detection event on a superconducting nanowire single-photon detector (SNSPD 1). \textbf{b,} Hybrid entanglement is generated in parallel with two OPOs, a single-mode squeezer (OPO I) for the CV component and a two-mode squeezer (OPO IIb) for the DV component, and is heralded by a detection event on SNSPD 2. The input DV qubit to be converted is then mixed with the DV mode of the hybrid entanglement and an original Bell-state measurement (BSM, blue panel) is performed. The conversion success is heralded by a detection event on SNSPD 3 and a quadrature conditioning on a homodyne detection (HD). The resulting output qubit is finally characterized by full quantum state tomography with a high-efficiency HD. Photodiodes P1, P2, and P3 are used for phase control and stabilization. \textbf{c,} Quantum circuit representation of the converter. Hybrid entanglement is prepared by applying a Hadamard gate (H) on the DV mode and mixing the DV and CV modes with a C-NOT gate. The input DV qubit is mixed with the DV mode of the entangled state by applying a C-NOT gate. The output of the conversion is conditioned on the BSM measurement outcome that heralds the teleported CV qubit.}
\label{fig2}
\end{figure*}

The experimental setup is detailed in Fig. \ref{fig2}, together with the associated quantum circuit. Three optical parametric oscillators (OPOs, see Methods) are simultaneously used to independently create the initial resource states, i.e., the input DV qubit to be converted and the optical hybrid entanglement that enables the teleportation process. They are pumped below threshold by a continuous-wave, frequency-doubled Nd:YAG laser. Their bandwidths are precisely matched to ensure high indistinguishability. The OPOs in the setup correspond to different stages of the converter: OPO IIa for DV qubit creation, and the pair of OPO IIb and OPO I for qubit conversion via hybrid entanglement. Each state preparation is heralded by a photon detection on high-efficiency superconducting nanowire single-photon detectors (SNSPDs), with a system detection efficiency above $85\%$ at 1064 nm and a dark count rate below 10 counts per second \cite{Le-Jeannic:2016:OptLett}.

To implement the conversion process, the first step is to create the initial DV qubits, as shown in Fig. \ref{fig2}a. For this purpose, we use a type-II phase-matched OPO (OPO IIa) that emits a two-mode squeezed vacuum state, i.e., correlated signal and idler photons. The OPO is used in the very-low pump regime. The idler photon is detected on a SNSPD after frequency filtering. This detection event heralds the generation of a high-purity single photon, with a heralding efficiency of $\eta=71.2\pm1.5\%$, given by the escape efficiency of the OPO. This state shows strong non-classical features, with a negativity of the Wigner function of $-2 \pi (0.45 \pm 0.02)$ and a second-order autocorrelation function at zero time delay of $g^{2}(0) = 0.10 \pm 0.03$. To create the DV qubits, i.e., superpositions of vacuum and single-photon states, we apply a controllable phase-sensitive displacement on the heralding mode, as realized in \cite{neergaard2010optical} for engineering CV qubits. This displacement is obtained by mixing the idler mode with an attenuated coherent state, making the photons from the two modes indistinguishable. Depending on the amplitude and phase of the coherent state, we can herald DV qubits in the Fock-sate basis $\{\ket{0}, \ket{1}\}$ of the form $\ket{\psi}_{DV}=c_{0}\ket{0}+c_{1}e^{i\theta}\ket{1}.$  The relative phase $\theta$ between the idler mode and the displacement mode is actively controlled with a digital lock (phase noise of about $10\%$). Details on the preparation of the DV qubits are given in the Supplementary Information. We generate a set of six qubits spanning the whole Bloch sphere. The states are created at a rate ranging from 250~kHz for the state $\ket{1}$ to 2~MHz for the state $\ket{0}$ with a rate of 500~kHz for the four balanced qubits along the sphere equator.

Concurrently, hybrid entanglement is created from the joint operation of a single-mode squeezer and of a two-mode squeezer, as sketched on Fig. \ref{fig2}b. The idler mode of OPO IIb is mixed with a small tapped fraction of the squeezed vacuum emitted by OPO I in an indistinguishable fashion. When balancing the counts from each OPO, a click on SNSPD 2 heralds a maximally entangled hybrid  state of the form $\ket{\psi} \propto \ket{0}\ket{cat-}+\ket{1}\ket{cat+}$, where $\ket{cat+}$ and $\ket{cat-}$ respectively denote even and odd Schr\"odinger cat states of amplitude $|\alpha|=0.9$. The hybrid entangled state is generated at a rate of about 400~kHz. Details on the engineering of hybrid entanglement have been reported elsewhere \cite{Huang:2019:NewJPhys, Morin:2014:NatPhot}. 

\begin{figure*}[t!]
\includegraphics[width=1.8\columnwidth]{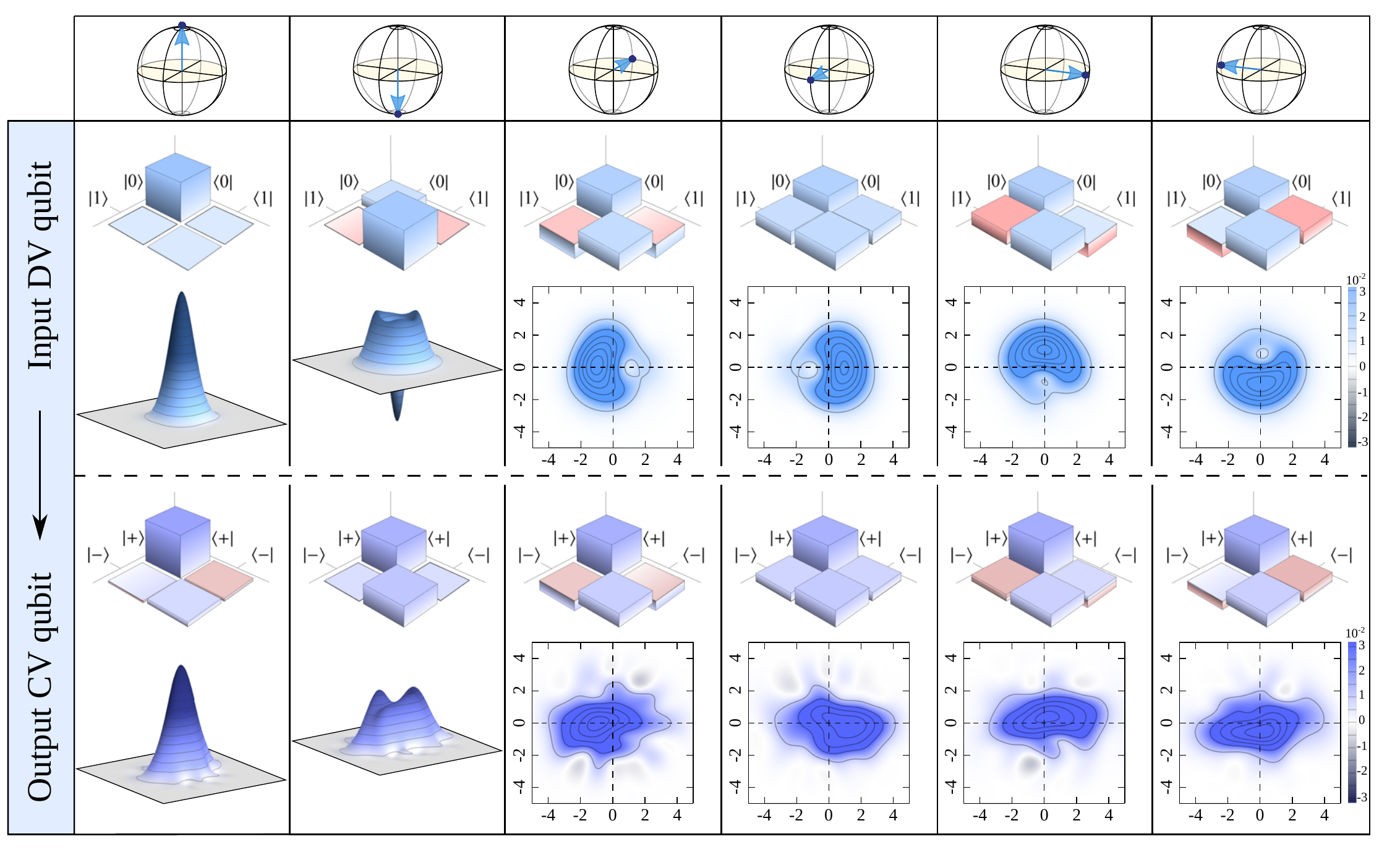}
\caption{\textbf{Qubit characterization before and after conversion.} A set of six qubits distributed on the Bloch sphere is converted from DV (top row) to CV encodings (bottom row). The density matrices of the input DV qubits are each reconstructed from 40000 quadrature measurements and projected onto the $\{\ket{0},\ket{1}\}$ Fock-state basis. The associated Wigner functions are also given. After conversion, the output qubits are reconstructed from 3691, 2185, 2802, 2657, 3641 and 2839 quadrature measurements respectively. The density matrices are projected onto the $\{\ket{cat+},\ket{cat-}\}$ cat-state basis, labeled $\{\ket{+},\ket{-}\}$ here. The real parts of the density matrices are shown in blue while the imaginary parts are shown in red. The Wigner functions are plotted in 3D for the phase-insensitive computational-basis elements, and are shown in 2D top-view for the others to provide a better visualization of the phase of the qubits located along the Bloch sphere equator.}
\label{fig3}
\end{figure*}

With both resources introduced, we now turn to the teleportation process. In order to realize the Bell-state measurement (BSM), the DV qubit is brought to interfere on a 50/50 beam splitter with the DV mode of the hybrid entangled state, with an interference visibility of about 98\%. A specific hybrid BSM is performed using a combination of photon counting and homodyne conditioning (see Supplementary Information). Specifically, we tap a very small fraction $R$ of the beam and send it to SNSPD 3 for heralding. After this first detection, the homodyne detection then functions as a parity check that verifies if the remaining state is indeed vacuum. This is realized by selecting the measured quadratures $q$ centered around $q=0$ in a given conditioning window $\Delta$. For our implementation we chose for the BSM parameters $R = 10\%$ and $\Delta = 0.5$ $\sigma_{0}$ where $\sigma_{0}$ is the standard deviation of the vacuum shot noise. These parameters ensure simultaneously an efficient measurement and a practicable operational rate.

Upon four successful events, i.e., detections on SNSPD 1, 2 and 3, and homodyne conditioning, the initial DV qubit is teleported onto a CV qubit of the form $\ket{\phi}~=~c_{0}\ket{cat+}+c_{1}e^{i\theta}\ket{cat-}$, thus completing the computational basis conversion from DV to CV. Details on the coincidental detections are given in the Methods. The overall rate of the protocol implementation is of about 3 events per minute. In the following, we benchmark the performance of the encoding conversion process.

\begin{figure}[t]
\vspace{-0.4cm}
\includegraphics[width=0.95\columnwidth]{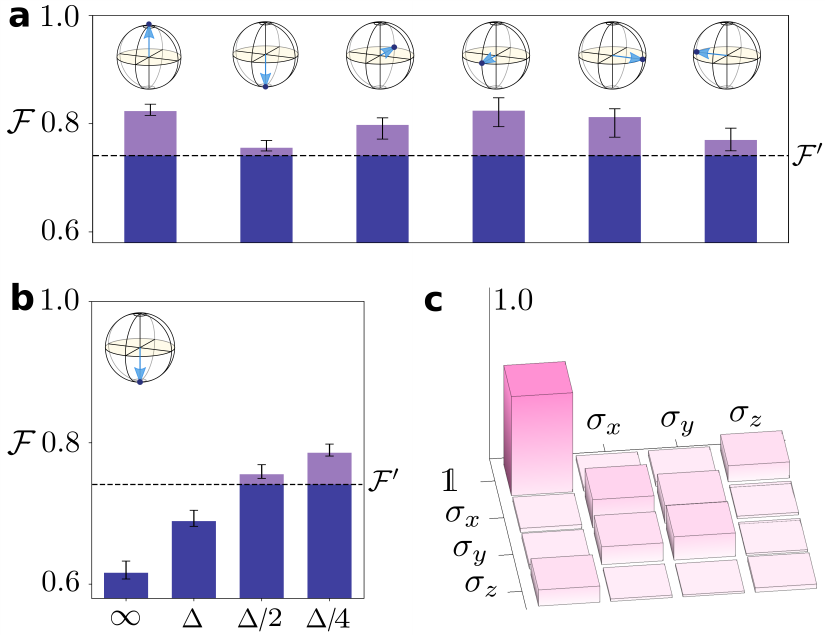}
\caption{\textbf{Characterization of the conversion process.} \textbf{a,} Experimental teleportation fidelity for the six input DV qubits projected into the $\{\ket{0},\ket{1}\}$ Fock-state basis and the corresponding teleported qubits projected into the $\{\ket{cat+},\ket{cat-}\}$ cat-state basis. The adapted classical limit of $\mathcal{F}^\prime=74.1\%$ is indicated by the dotted line and the average fidelity is of $\overline{\mathcal{F}} = 79.7\%$. The error bars take into account the state preparation (see Supplementary Information). \textbf{b,} Fidelity for the conversion of the single-photon input as a function of the homodyne conditioning window width. $\infty$ denotes the absence of conditioning and $\Delta=\sigma_0$ is equal to the standard deviation of the vacuum shot noise. \textbf{c,} Absolute values of the process matrix $\chi$ reconstructed using convex programming. The first element $\chi_{11}$ corresponds to the identity and reaches a value of 0.58 above the 1/2 classical limit.}
\label{fig4}
\end{figure}

First, we characterize the input DV qubits by full quantum state tomography performed with high-efficiency homodyne detection. The states are reconstructed via a maximum-likelihood algorithm \cite{Lvovsky} and only corrected for 18$\%$ detection losses. The density matrices and Wigner functions of the six initial DV qubits are shown in Fig. \ref{fig3}. We calculate the fidelity of the input states with the ideal pure targeted states and obtain a value of $71.2\%\pm1.5\%$ for the input single photon and an average fidelity of $79.3\%\pm1.9\%$ over the full Bloch sphere. The fidelity is limited mostly by the initial heralding efficiency $\eta$ given by the OPO escape efficiency and by the phase noise of the lock performed to define the qubit phase. 

Next, we study the teleported states. Each output CV state is also reconstructed by full quantum state tomography in a Hilbert space of dimension 10. About 2500 coincidental events are used for each reconstruction, with a detection loss of $15\%$ taken into account. Figure \ref{fig3} shows the density matrices projected onto the cat-state basis $\{\ket{cat+},\ket{cat-}\}$ and the Wigner functions of the teleported states. We observe that the off-diagonal coherence terms of the qubits are retained through the process, meaning that the original phase information of the qubit is well preserved via the conversion. The reduction in the amplitude of these terms mainly originates from the overall phase noise of the implementation for which 14 active locks are operated in parallel (see Methods). The projections of the Wigner functions provide another representation of this phase preservation, a strong signature of the efficiency of the process.

To assess the performance of the converter, we calculate the fidelity between the input qubit $\hat{\rho}_{in, DV}$ projected onto the DV basis and the output qubit $\hat{\rho}_{out, CV}$ projected onto the CV basis. The fidelity is given by $\mathcal{F}~=~\Big(\textrm{Tr} \sqrt{\sqrt{\hat{\rho}_{in, DV}}\hat{\rho}_{out, CV}\sqrt{\hat{\rho}_{in, DV}}} \Big)^{2}$  \cite{Jozsa1994}. The success of the conversion process, namely its quantum character, is assessed comparing the fidelity to a classical bound, i.e., obtained without the use of entanglement. The usual bound of $\mathcal{F} = 2/3$ \cite{OldF} can only be applied to postselected process with pure input qubits and mixed output states. However, this classical bound strongly increases when mixed input states are considered. Following a methodology defined in \cite{newBound}, we adapt this benchmark to our mixed input states (see Supplementary Information) and calculate the classical bound to be $\mathcal{F}^\prime = 74.1 \%$. As shown in Fig. \ref{fig4}a, it is exceeded over the full Bloch sphere, with an average fidelity of $\overline{\mathcal{F}}_{exp} = 79.7^{+0.7}_{-1.0}\,\%$. This value confirms the success of the encoding conversion. Moreover, we stress that the fidelities are here not conditioned on a successful detection after the conversion and the process is thereby free of any postselection.

As can be seen in Fig. \ref{fig4}a, the teleportation fidelity decreases as the single-photon component of the input qubit increases. Indeed, the ensuing contribution of the multi-photon events leading to unwanted coincidental events results in the addition of vacuum in the measured teleported state. This effect is mitigated first by the low-reflectivity of the BSM tapping ratio. In addition, homodyne conditioning on quadrature values around $q=0$ drastically improves the efficiency of the process. Figure~\ref{fig4}b illustrates the effect of the conditioning window width for the hardest conversion process, namely the teleportation of the single-photon state to the odd cat state. The process cannot exceed the classical bound if this conditioning step is discarded and the fidelity increases as the window is narrowed, demonstrating the efficiency of the hybrid BSM combining photon counting and homodyne conditioning. The multi-photon contribution would be completely eliminated should photon-number resolving detectors be used, yet at the expense of a reduced operational rate. Finally, the fidelity is limited by the intrinsic fidelity between the initial state at the output of OPO~I and the targeted cat states, which is of about 85$\%$ for the pump power used in our experiment. 

To further investigate the conversion process, we also characterize it by means of quantum process tomography. Starting from four initial qubits, quantum process tomography provides the process matrix $\chi$ that fully defines the map from an arbitrary input state to the resulting output state in the basis of the Pauli operators $\{\mathbb{1},\sigma_x,\sigma_y,\sigma_z\}$ \cite{Riebe2007}. The first term of the matrix $\chi_{11}$ gives the process fidelity while other terms quantify potential errors. We performed such matrix reconstruction via convex programming (see Supplementary Information). Figure \ref{fig4}b shows the reconstructed matrix, with $\chi_{11}=(58 \pm 3)\%$, above the $1/2$ classical threshold.

Our demonstration was made possible by the combination of high-purity and indistinguishable sources, active phase stabilization of multiple paths, a low-loss implementation and the use of highly-efficient single-photon and homodyne detectors. However, as heralded non-Gaussian states are used concurrently, the rate of the protocol exponentially decreases with the number of modes, a roadblock generally encountered in distributed quantum computing or long-distance quantum communication where resources have to be combined. Several routes can tackle this inefficiency. Firstly, a technique would be to assist the heralded sources with quantum memories to enable the synchronization of the stochastic events, making the overall success rate decreasing then only polynomially with the number of modes. Such memories have been implemented either all-optically \cite{Yoshikawa2013, Bouillard2019} in the context of quantum state engineering or by interfacing the optical modes with matter systems in very efficient manners for quantum networks \cite{Wang2019,hoffet2020efficient}. These techniques could also facilitate the use of larger amplitude optical cat states, as demonstrated by direct generation \cite{Huangcat} or breeding procedures \cite{breeding}. Secondly, we also emphasize that our input-output circuit is compatible with on-demand sources based on solid-state emitters that have also recently been used in teleportation protocols \cite{Reindl2018, BassoBasset2019}, promising much higher rates in the future. Finally, bit flip on the DV mode encoded in the Fock-state basis naturally occurs upon propagation of the states in lossy quantum channels. Extending hybrid entanglement, and so qubit conversion, to more loss-resilient DV encodings is also a necessary route to follow. Several non-postselected schemes have been proposed either with polarization \cite{Kwon2015} or time-bin qubits \cite{Gouzien2020}. In this regards, we note that the conversion fidelity achieved for our input single-photon state beats the stringent non-post-selected fidelity bound defined for such dual-rail qubits (See Supplementary Information) \cite{Furusawa}. 

In conclusion, we experimentally realized an optical qubit converter, enabling the faithful conversion of quantum information from discrete- to continuous-variable qubits. The classical limit of conversion is exceeded over the full Bloch sphere, with an average fidelity above 79\%. Our realization involves independent preparations of the resources, with an incoming discrete single-photon qubit and an output freely-propagating cat qubit ready for operations. This work represents a significant advance towards connecting disparate quantum hardwares and processing schemes, a grand challenge to scale up quantum technology infrastructures.

\section*{\textbf{Methods}}
\small{

\noindent{\bf Optical parametric oscillators.}
Three optical parametric oscillators (OPOs) are used in parallel, with a bandwidth of $50$~MHz and a free spectral range of $4.3$~GHz. They are pumped below threshold by a frequency-doubled continuous-wave laser (Innolight GmbH, Diabolo) at $532$~nm. All OPOs are resonant for all fields (pump, signal and idler) simultaneously. To create the hybrid entangled state, two semi-monolithic linear-cavity OPOs are used where the input mirrors are coated directly on the crystal (coating by Layertech GmbH, high reflectivity at $1064$~nm and $95\%$ reflectivity at 532 nm) and the output mirrors have a radius of curvature of $38$~mm (coating by Layertech GmbH, with $90\%$ reflectivity at $1064$~nm and high reflectivity at $532$~nm). Single-mode squeezed vacuum is created by OPO I ($10$~mm type-I phase-matching PPKTP crystal from Raicol, pumped at $15$~mW with $4.5$~dB of output squeezing), and $7$~$\%$ is tapped for heralding the creation of a photon-subtracted squeezed vacuum. Resonance is achieved via tuning the cavity length and crystal temperature. The count rate of the OPO I heralding is balanced with the one of the idler of the two-mode squeezed vacuum generated by OPO IIb, with a heralded single-photon purity of $80\%$ ($10$~mm type-II phase-matching KTP crystal from Raicol, pumped at $3.5$~mW). To enable triple-resonance, the cavity length, crystal temperature and laser wavelength are all tuned. For the DV qubit creation a second two-mode squeezer is used (OPO IIa, $10$~mm type-II KTP crystal from Raicol, pumped at $2$~mW). To achieve simultaneous triple-resonance of both OPO IIa and OPO IIb we tune the angle of the OPO IIa crystal as an additional free parameter to compensate for the loss of the wavelength degree of freedom, already engaged for the lock of OPO IIa. This requires a different cavity design, in which the free standing input mirror forms a straight-cut linear cavity (input and output coupler reflectivities remain the same). This stable triple resonance of two type-II phase-matched OPOs pumped with the same laser is non-trivial and, to the best of our knowledge, achieved here for the first time.

\vspace{10pt}

\noindent{\bf SNSPD operation and coincidence detection.}
Three superconducting nanowire single-photon detectors are operated at 1.3 K in a custom-made cryocooler (MyCryoFirm). Before arriving on each detector, the heralding photons are filtered by a broadband interference filter (Barr Associates, bandwidth of $125$~GHz) and a homemade high-finesse Fabry-Perot micro-cavity (free spectral range $330$~GHz, bandwidth $320$~MHz). Three heralding events (SNSPD 1 for input qubit creation, SNSPD 2 for hybrid entanglement generation, and SNSPD 3 for Bell-state measurement) need to be detected in a specific coincidence window to herald the success of the teleportation. The three-fold-coincidence-detection event is analyzed by a programmable ultra-fast coincidence detection module (ID Quantique, ID900 Time Controller) that defines the three-fold coincidence from a series of two two-fold coincidences. First, the coincidence between SNSPD 1 and SNSPD 2 is assessed, and the result of this coincidence detection is coincidentally detected with SNSPD 3. Upon such an event two homodyne detections corresponding to the BSM and the measurement of the teleported state are triggered. To ensure high fidelity, each two-fold coincidence window is set to $1.5$~ns, which ensures a temporal mode overlap of $86 \%$ between all three photons to be detected. With this coincidental window the overall experiment rate is of about 12 events per minute, which is the minimum operational rate that is allowed while ensuring passive phase preservation between two consecutive events. This rate reduces to 3 events per minute when applying the additional homodyne conditioning.

\vspace{10pt}

\noindent{\bf Phase control and stabilization.}
Fourteen locks control various phases in parallel and are actively monitored during each experimental run of $10$-$16$~hours. To be resonant with the pump field, each OPO is locked via the Pound-Drever-Hall method. The infrared laser output at $1064$~nm, filtered with a mode-cleaner cavity that is locked via tilt-locking, is used as a seed to monitor the resonance conditions of the OPOs and to lock digitally the relative phase between the pump and the down-converted fields on the maximum of amplification (ADuC7020 Analog Devices). The relative phases between (1) the heralding mode of OPO IIa and the displacement for the qubit creation, (2) the heralding modes of OPO I and OPO IIb for the entanglement generation, and (3) the two DV modes mixed before the BSM, are locked digitally in the same manner. The three high-finesse filtering micro-cavities are locked with a two-stage digital lock. The seed beams are additionally used for phase calibration of the homodyne detections. To ensure phase stability, a sample-and-hold procedure is employed. The seed beams are used to readjust all locks during a $50$~ms period, apart from the mode-cleaner cavity and the OPO pump locks that are continuously on. The following $50$~ms window is used for data acquisition, with the seed beams turned off and SNSPDs paths unblocked.
}

\noindent \textbf{Acknowledgements} This work was supported by the QuantERA grant ShoQC, by the French National Research Agency (HyLight project ANR-17-CE30-0006, ShoQC project 19-QUAN-0005-05), and by the Horizon 2020 Research and Innovation Programme via the Twinning project NonGauss (951737). G.G. was supported by the European Union (Marie Curie Fellowship HELIOS IF-749213) and T.D. by Region Ile-de-France in the framework of DIM SIRTEQ. J.L. is a member of the Institut Universitaire de France. \\

\noindent\textbf{Author contributions}
T.D. and B.A. performed the experiment, developed implementation techniques and analyzed the data. G.G., A.C., and H.L.J.  contributed to the preparation of the setup. J.L. designed the research and supervised the project. All authors discussed the results and contributed to the writing of the manuscript.\\

\clearpage
\renewcommand{\thefigure}{S\arabic{figure}}
\renewcommand{\thetable}{S\arabic{table}}
\setcounter{figure}{0}

\onecolumngrid

\centerline{\textbf{SUPPLEMENTARY INFORMATION}}

\section{Preparation of the discrete-variable input qubits}

\begin{figure}[!b]
\centerline{\includegraphics[width=0.7\columnwidth]{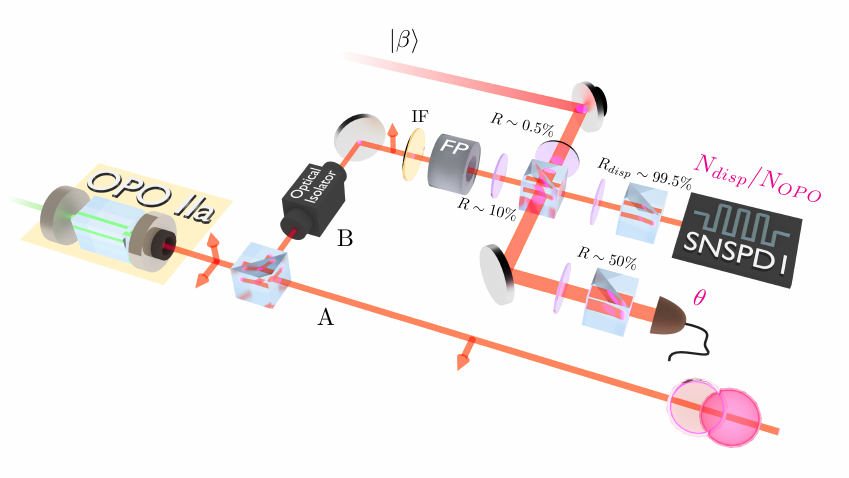}}
\vspace{-0.5cm}
\caption{\textbf{Experimental setup for the discrete-variable qubit preparation}. The heralding mode of a type-II optical parametric oscillator is combined with an attenuated coherent state $\ket{\beta}$ on an unbalanced polarizing beamsplitter. The beamsplitter is set to get $90\%$ of the detection events on the SNSPD from the OPO, and only $0.5\%$ from the coherent state. The other port of the beamsplitter is sent to a photodiode used to lock the relative phase between the two beams and thus the DV qubit phase $\theta$. The interference fringe for the lock is obtained by projecting the two orthogonal polarizations with a half-wave plate at $45^{\circ}$ before a second polarizing beamsplitter. The coherent state is attenuated a second time before the SNSPD by setting the transmission of the displacement polarization at $0.5\%$ with a third PBS. Once the half-wave plates are set, the amplitude of the coherent state $\beta$ is tuned upstream with a variable neutral-density filter. The setting is done at the single-photon level by comparing the OPO rate $N_{OPO}$ and the displacement rate $N_{disp}$ on the SNSPD. The heralding photons from the OPO are frequency filtered by an interferential filter (IF) and a home-made Fabry-Perot cavity (FP). Upon detection on the SNSPD, a discrete-variable qubit $\ket{\psi}_{A}=c_{0}\ket{0}_{A}+c_{1}e^{i\theta}\ket{1}_{A}$ with $c_{0}=\sqrt{N_{disp}/(N_{disp}+N_{OPO})}$ is generated.}
	\label{Qubit_preparation}
\end{figure}

The preparation of the discrete-variable qubits encoded in the Fock-state basis $\{\ket{0},\ket{1}\}$, $\ket{\psi}_{A} = c_{0} \ket{0} + c_{1}e^{i\theta}\ket{1}$, is based on a two-mode squeezed vacuum state emitted by a type-II phase-matched optical parametric oscillator operated far below threshold. The scheme is given in Fig. \ref{Qubit_preparation}. Signal mode $A$ and idler mode $B$ are separated by a polarizing beamsplitter and, after displacement, mode $B$ is sent onto a superconducting nanowire single-photon detector (SNSPD) for heralding.

In the ideal case, the two-mode squeezed vacuum at the output of the OPO can be written as:
\begin{equation}
\ket{\psi}_{AB} \propto \ket{0}_{A}\ket{0}_{B} + \lambda \ket{1}_{A}\ket{1}_{B} + \mathcal{O}(\lambda)
\end{equation}
where $\lambda$ is related to the squeezing parameter $s=(1-\lambda)/(1+\lambda)$. When a small displacement is applied to mode $B$, $\hat{D}_{B}(\beta) \sim \mathds{1} + \beta \hat{b}^{\dagger} - \beta^{*} \hat{b}$, the two-mode state becomes:
\begin{equation}
\ket{\psi}_{AB} \propto (\beta \ket{0}_{A} + \lambda \ket{1}_{A})\ket{1}_{B} + \lambda \beta \ket{1}_{A}\ket{2}_{B}.
\end{equation}
In the ideal case of a $\ket{1}\bra{1}_{B}$ projection, a detection on the SNSPD heralds the preparation of the DV qubit:
\begin{equation}
\ket{\psi}_{A} \propto \beta \ket{0}_{A} + \lambda \ket{1}_{A}.
\end{equation}
This coherent superposition originates from the two indistinguishable events detected on the SNSPD, i.e., either a click from the displacement mode heralding then the state $\ket{0}_{A}$ or a click from the OPO mode heralding then the state $\ket{1}_{A}$. The superposition weight and phase are controlled by the amplitude and phase of the displacement. The superposition weight is defined by the ratio of heralding rates originating from the OPO, labeled $N_{OPO}$, and from the displacement beam, $N_{disp}$. They are linked to parameters $\lambda$ and $\beta$ by the relation:
\begin{equation}
\frac{\beta}{\lambda}=\sqrt{\frac{N_{disp}}{N_{OPO}}}.
\end{equation}
Overall, the discrete-variable qubit can be represented on a Bloch sphere with the angles $\phi$ and $\theta$ defined by:
\begin{equation}
c_{0}=\cos(\phi/2)=\sqrt{\frac{N_{disp}}{N_{disp}+N_{OPO}}}, \quad \textrm{and} \qquad \theta=\pi-arg(\beta).
\end{equation}

As shown on Fig. \ref{Qubit_preparation}, the idler photons are first frequency filtered with an interferential filter (Barr Associates, bandwidth $125$~GHz) and a home-made Fabry-Perot cavity (free-spectral range 330~GHz, bandwidth 320~MHz). The displacement is realized by mixing this mode with an attenuated coherent state $\ket{\beta}$ on an unbalanced polarizing beamsplitter. One port of the beamsplitter is sent to a SNSPD for heralding, while the other port is sent to a photodiode to lock the relative phase between the two paths, thus defining the qubit phase $\theta$. About 10$\%$ of the heralding path of the OPO is reflected towards the photodiode, while only 0.5$\%$ of the attenuated coherent state is sent to the SNSPD. After the beamsplitter, the beams copropagate in the same spatial modes but with orthogonal polarizations. To lock the relative phase the two polarizations are projected onto a second PBS with a half-wave plate set at $45^{\circ}$. Since the SNSPDs are polarization independent, it is not required to project the two polarizations before the detection. An additional PBS is still used to attenuate the coherent state that would otherwise saturate the SNSPD. This tunable beamsplitter is set to have a transmission of 0.5$\%$ for the polarization corresponding to the displacement beam. In the end once the polarization setting is performed, the half-wave plates are not rotated anymore, and the coherent state is attenuated before the mixing with a variable neutral-density filter.

With this experimental setting, arbitrary DV qubits can be prepared by tuning the displacement amplitude and phase. Without any displacement, the heralded state in mode $A$ is a single photon $\ket{1}_{A}$. For an \textit{infinitely} large displacement - or equivalently when the OPO path is closed -, the vacuum $\ket{0}_{A}$ is heralded. Equally-weigthed qubits are prepared by tuning the variable ND filter to equalize both rates directly at the single-photon level. The phase of the qubit is then defined by the locking point on the fringe. This lock is performed digitally with a microcontroller (Analog Devices Aduc 7020) \cite{Huang2014}. The heralded DV qubit in mode $A$ is then available for any subsequent operation. It can be characterized using state tomography, or used as an input of the conversion protocol presented in this work.  

\section{Full experimental setup}
\label{FES}

\begin{figure}[!t]
\includegraphics[width=0.88\textwidth]{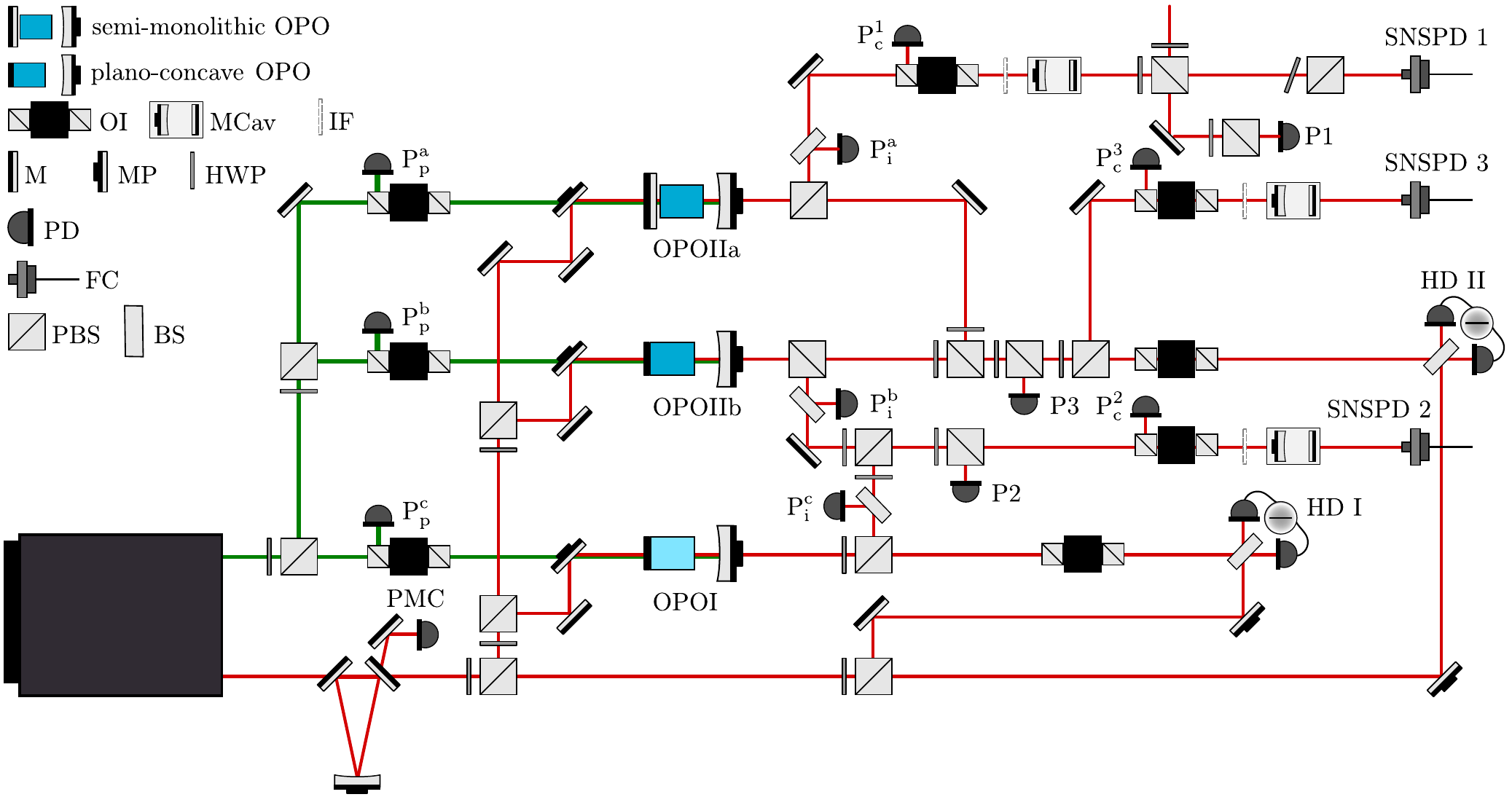}
\caption{\textbf{Full experimental setup.} OI Optical Isolator, MCav homemade micro-cavity, IF Interference Filter, M Mirror, MP Mirror mounted on a piezo-electric transducer, PD Photodiode, FC Fibercoupler leading to the superconducting nanowire single-photon detectors SNSPDs, BS Beamsplitter, PBS Polarizing beam splitter, HWP Half-wave plate, HD Homodyne detection. Details are given in the text. Some waveplates and lenses are not shown for the sake of readability.}
\label{ExpS}
\end{figure}

We now turn to the detailed experimental setup, as illustrated in Fig. \ref{ExpS}. We start from a continuous-wave frequency-doubled Nd:YAG laser (Diabolo, Innolight GmbH), providing 250 mW light output at 1064 nm and 500 mW at 532 nm. To reduce intensity noise and clean the transverse mode profile, the infrared output is filtered by a triangular filtering cavity (finesse 3500, transmission $80\%$), which is locked via tilt locking.

The three triply-resonant OPO cavities are locked on resonance with the pump by the Pound-Drever-Hall technique (12 MHz modulation) on the back-reflection paths of optical isolators, which are detected by photodiodes $\textrm{P}_{\textrm{p}}^{\textrm{a}}$, $\textrm{P}_{\textrm{p}}^{\textrm{b}}$ and $\textrm{P}_{\textrm{p}}^{\textrm{c}}$. Resonances of the OPO cavities with the IR modes are monitored via photodiodes $\textrm{P}_{\textrm{i}}^{\textrm{a}}$, $\textrm{P}_{\textrm{i}}^{\textrm{b}}$, $\textrm{P}_{\textrm{i}}^{\textrm{c}}$. The OPO locks are active during the whole experiment. In contrast, nine other locks are based on seed beams that are present in the framework of a sample-and-hold procedure during a $50$~ms period used for readjustment. The following $50$~ms window is used for data acquisition, meaning the seed beams are off while the SNSPDs are unblocked.

On the conditioning paths, as presented before for the qubit generation, interference filters (Barr Associates, bandwidth $125$~GHz) and  homemade high-finesse Fabry-Perot cavities (FSR $330$~GHz, bandwidth $320$~MHz) are used to ensure the heralding of the degenerate-frequency component. These cavities are locked using counter propagating injected seed beams, which are eliminated by optical isolators and detected on $\textrm{P}_{\textrm{c}}^{\textrm{1}}$, $\textrm{P}_{\textrm{c}}^{\textrm{2}}$, $\textrm{P}_{\textrm{c}}^{\textrm{3}}$ via a high-finesse maximum-amplitude search algorithm on microcontrollers \cite{Huang2014}.

Hybrid entanglement is created by mixing the heralding paths of OPO IIb (idler mode) and OPO I ($7$~$\%$ are tapped) in an indistinguishable fashion on a polarizing beam splitter \cite{Morin2014}. In this way the origin of the photon upon detection on SNSPD 2 cannot be inferred and therefore the heralding event leads to the state proportional to  $a\ket{0}\ket{cat-}+e^{i\psi}b \ket{1}\ket{cat+}$, where the weights $a$ and $b$ are controlled by the respective count rates of the OPOs ($a$ and $b$ are equal in this work) and the phase $\psi$ is measured on $\textrm{P}2$ and locked via a microcontroller running a maximum-amplitude algorithm search.The qubit phase is locked in the same manner via $\textrm{P}1$. As this algorithm only locks on the maximum of the fringe, a tilted $\lambda/2$-plate allows us to tune the relative path length as the heralding OPO beam and the attenuated displacement beam have orthogonal polarizations on their path towards SNSPD 1. Similarly, the relative phase between the qubit and the DV component of the hybrid entanglement is locked via a microcontroller on $\textrm{P}3$.

\section{Conversion protocol: comparison with theory}

In the following, we model the conversion protocol taking into account all experimental imperfections, exclusive of phase noise, and compare our experimental results to the expected states.

The heralded DV qubit is composed of a qubit fraction corresponding to the heralding efficiency $\eta_{qubit}$ of a single photon generated by the OPO. The state can be written as:
\begin{equation}
\hat{\rho}_{A} = (1-\eta_{qubit})\ket{0}\bra{0}_{A} + \eta_{qubit}\ket{Q}\bra{Q}_{A}
\end{equation}
\noindent where $\ket{Q}$ represents the target input DV qubit $\ket{Q}=c_{0}\ket{0}_{A}+e^{i\theta}\ket{1}_{A}$.

The hybrid entangled state is written in the experimental basis $\{\ket{0},\ket{1}\}$ for mode $B$ and $\{\hat{S}(r)\ket{0},\hat{a}\hat{S}(r)\ket{0}\}$ for mode $C$ for the DV and CV basis respectively, where $\hat{S}$ and $\hat{a}$ correspond to the squeezing and annihilation operators respectively. Similarly to the input qubit, the experimental hybrid entangled state is simulated by taking into account the respective heralding efficiencies $\eta_{DV}$ and $\eta_{CV}$ of the DV and CV modes. 

The modes $A$ and $B$ are then mixed on a 50/50 beamsplitter. Mode $A$ is traced out and mode $B$ will be subject to the Bell-state measurement (BSM). First, a small fraction of the light is tapped with a beamsplitter of reflectivity $R$ towards a single-photon detector located in mode $D$. The transmission up to the SNSPD is labeled $\eta_{SNSPD}$. The bucket detection is then described by the operator $\hat{\prod}_{D}=\sum_{n=1}^{2}[1-(1-\eta_{SNSPD})^{n}]\ket{n}\bra{n}_{D}$.

The untapped mode is sent to a homodyne detection in mode $B$ for completing the BSM. Homodyne conditioning around $q=0$ in a conditioning window $\Delta$ is described by the transformation
\begin{equation}
\hat{\rho} \rightarrow \hat{\rho}'=\textrm{Tr}_{B}(\int_{-\Delta/2}^{+\Delta/2}dq\ket{q}\bra{q}_{B}\hat{\rho})
\end{equation}
\noindent with $\ket{q}$ the eigenvector of the quadrature operator $\hat{q}_{\theta}=\hat{X}cos\theta+\hat{P}sin\theta$ where $\hat{X}$ and $\hat{P}$ denote the canonical position and momentum observables of the measured field and $\theta$ is the phase of the local oscillator. The efficiency of the homodyne detection $\eta_{HD}$ is finally taken into account, and amounts to the transformation 
\begin{equation}
A_{00}^{\Delta} \rightarrow A_{00}^{\Delta}, \hspace{1cm} A_{11}^{\Delta} \rightarrow \eta_{HD} A_{11}^{\Delta} + (1-\eta_{HD})A_{00}^{\Delta}, 
\end{equation}
\noindent with $A_{ij}^{\Delta}=\int_{-\Delta/2}^{+\Delta/2}dx\ket{x}\bra{x}_{B}\ket{i}\bra{j}_{B}$ and noticing $A_{01}^{\Delta} = A_{10}^{\Delta} = 0$ for all $\Delta$.

Finally we also consider the mode matching $\chi$ between the DV modes before the BSM. It is composed of a temporal part $\chi_{temp}$ that originates from the finite temporal coincidence window and of a spatial part $\chi_{spa}$ that corresponds to the visibility between the two DV modes. The non-perfect mode matching is then directly equivalent to losses seen by the homodyne detection, that transform the efficiency by:
\begin{equation}
\eta_{HD} \rightarrow \eta_{HD}'= \chi_{temp}.\chi_{spa}.\eta_{HD}.
\end{equation}

\begin{table}[b!]
    \centering
    \renewcommand{\arraystretch}{1.3}
\begin{tabular}{|*{11}{c|}}
  \hline
  Parameter & $\eta_{qubit}$ & $\eta_{DV}$ & $\eta_{CV}$ & $S(r)$ & $R$ & $\eta_{SNSPD}$ & $\Delta$ & $\eta_{HD}$ &$\chi_{temp}$ & $\chi_{spa}$\\
  \hline
  Value & 0.72 & 0.85 & 0.85 & 4dB & 0.1 & 0.3 & 0.5$\sigma_{0}$ & 0.83 & 0.86 & 0.98\\
  \hline
\end{tabular}

\begin{tabular}{|*{11}{c|}} 
  \hline
  Input state & $\ket{1}$ & $\ket{0}$ & $\ket{0}-\ket{1}$ & $\ket{0}+\ket{1}$ & $\ket{0}-i\ket{1}$ & $\ket{0}+i\ket{1}$\\
  \hline
  Fidelity & 96.1$\%$ & 95.3$\%$ & 90.8$\%$ & 93.2$\%$ & 91.5$\%$ & 93.2$\%$ \\
  \hline
\end{tabular}
\caption{\textbf{Simulation parameters and fidelities.} The two tables provide the parameters for the simulation of the experiment and the fidelities between the simulated states and the measured ones. All the parameters are measured independently. }
\label{Tab:Simul}
\end{table}

\begin{figure*}[t!]
\includegraphics[width=0.85\columnwidth]{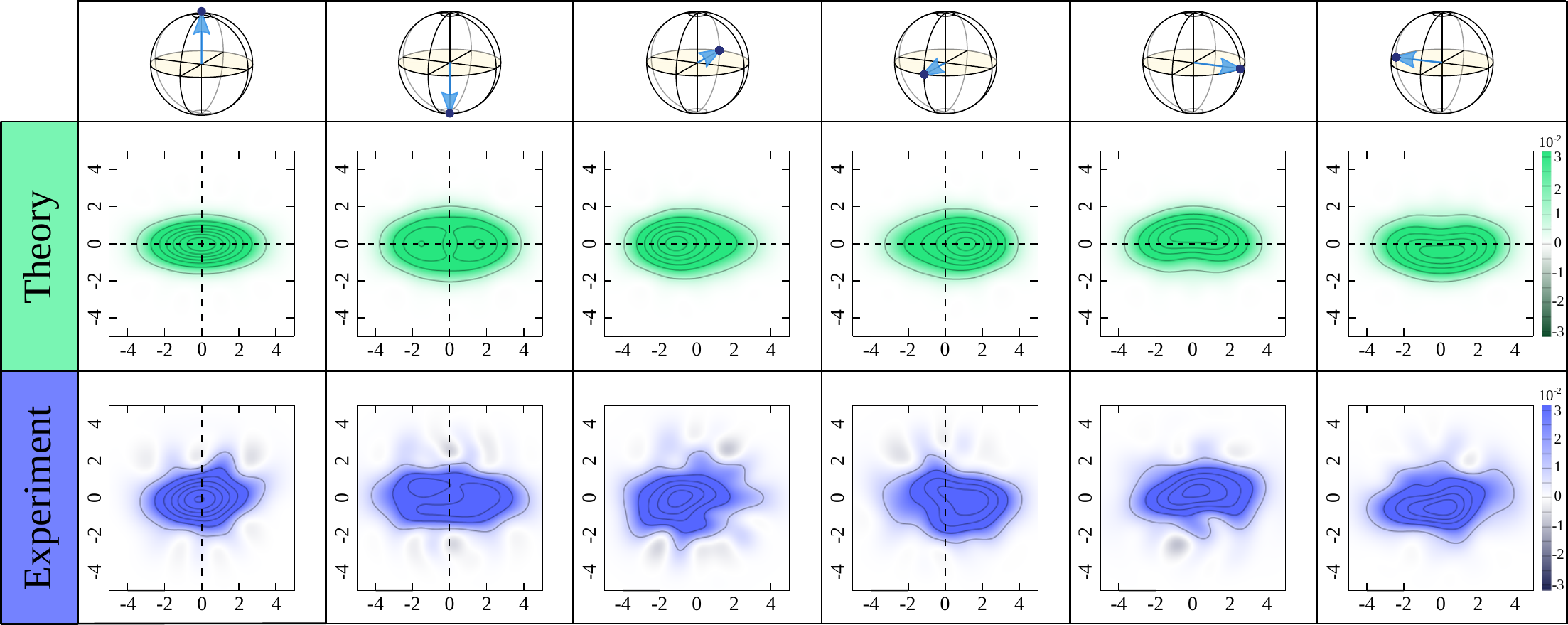}
\caption{\textbf{Simulated and experimental output states.} The Wigner functions of the six converted qubits  are plotted in 2D top-view to stress the qubit phase both in theory and experiment. The average fidelity is of about 93\%. }
\label{fig4}
\end{figure*}

With that model we can simulate the conversion protocol in our experimental configuration. Each parameter is measured independently and no free parameter is used to fit our result. Figure \ref{fig4} provides the Wigner functions of the simulated states and of our experimental states. The parameters and fidelities are given in Table \ref{Tab:Simul}, with an average fidelity between them of 93$\%$. The remaining percents originate from the phase noise and from the imperfect projection of the teleported state to the target squeezing.

\section{Error estimation on the fidelities}
\label{ErrorEst}

We detail our error estimation on the reported teleportation fidelities. The error bars come from the reconstruction uncertainties and the dispersion of possible input states per qubit. In the following we estimate the latter and most dominant source. 

Each measurement point takes approximately one day of data taking during which the input state cannot be measured. This leads to the necessity to pre- or post-measure the experimental input state, resulting in variations of the states that can be used to calculate the fidelity. Of course this value has to be considered carefully as it can lead to an overestimation of the performances of the protocol. We will estimate the minimal and maximal fidelities for each of our six experimental output states. We will eventually choose as the experimental input state for the fidelity calculation a state that represents the average input state our experiment provides. 

\begin{table}[b!]
\centering
\renewcommand{\arraystretch}{1.3}
 \begin{tabular}{|p{25mm}| p{25mm}| p{25mm}| p{25mm}| p{25mm}|} 
 \hline
 $\ket{Q}$ &$\ket{0}\bra{0}$ & $\ket{1}\bra{1}$ & $\ket{0}\bra{1}$ & $\ket{2}\bra{2}$ \\ [0.5ex] 
 \hline\hline
 $\ket{0}$ & $0.980 \pm 0.007$ & $0.0150 \pm 0.007$ &$0.000 \pm 0.000$ &  $0.005 \pm 0.00$ \\ 
 \hline
 $\ket{1}$ & $0.250 \pm 0.020$ & $0.710 \pm 0.020$ &$0.000 \pm 0.000$ &  $0.025 \pm 0.020$ \\
 \hline
 $\frac{1}{\sqrt{2}}(\ket{0} + e^{i\phi} \ket{1})$ & $0.615 \pm 0.021$ &$0.355 \pm 0.019$ & $(0.276 \pm 0.039  \pm 0.012e^{i\pi/2})e^{i\phi}$ &  $0.014 \pm 0.005$ \\ 
 \hline
\end{tabular} 
\caption{\textbf{Average input density matrix elements and standard deviations.}}
 \label{T1}
\end{table}

\begin{table}[b!]
\centering
\renewcommand{\arraystretch}{1.3}
 \begin{tabular}{|p{25mm}| p{12mm}| p{12mm}| p{12mm}|} 
 \hline
 $\ket{Q}$ &$c_{vac}$ & $c_{sp}$  & $c_{tp}$ \\ [0.5ex] 
 \hline\hline
 $\ket{0}$ & $0.980$ & $0.0150$ &  $0.005 $ \\ 
 \hline
 $\ket{1}$ & $0.250$ & $0.710$ &  $0.025$ \\
 \hline
 $\frac{1}{\sqrt{2}}(\ket{0} + e^{i\phi} \ket{1})$ & $0.260$ & $0.710$ & $0.014$ \\ [1ex]  \hline
\end{tabular} 
\caption{\textbf{Estimated vacuum, single-photon and two-photon components for the six input states.}}
 \label{T2}
\end{table}

\begin{table}[b!]
\centering
\renewcommand{\arraystretch}{1.3}
 \begin{tabular}{|p{10mm}| p{20mm}| p{20mm}| p{20mm}| p{20mm}| p{20mm}| p{20mm}|} 
 \hline
 $\ket{Q}$ & $\ket{0}$ &$\ket{1}$ & $\frac{1}{\sqrt{2}}(\ket{0}-\ket{1})$ & $\frac{1}{\sqrt{2}}(\ket{0}+\ket{1})$ & $\frac{1}{\sqrt{2}}(\ket{0}+i\ket{1})$ & $\frac{1}{\sqrt{2}}(\ket{0}-i\ket{1})$ \\ [0.5ex] 
 \hline\hline
 $F_{min}$ & 0.815 & 0.750 & 0.772 & 0.795 & 0.775 & 0.750 \\ 
 \hline
 $F_{max}$ & 0.836 &  0.769 & 0.811 & 0.848 & 0.828 & 0.792 \\
 \hline
 $F_{exp}$ & $0.823_{-0.008 }^{+0.013}$ & $0.756_{-0.006}^{+0.013}$ & $0.798_{-0.026}^{+0.013}$   &$0.824_{-0.030}^{+0.024 }$ & $0.812_{-0.037}^{+0.016}$ & $0.770_{-0.020}^{+0.022}$ \\
 \hline
\end{tabular}
\caption{\textbf{Error estimation on the fidelity considering the possible range of theoretical input states and each experimental output state.}}
\label{T3}
\end{table}

To this end, we consider a large set of experimental input qubits measured on several days and calculate the average value and standard deviation for each relevant density matrix element, as shown in table \ref{T1}. We then simulate for each input qubit all physical combinations of density matrix values. The matrix elements can be computed as
\begin{equation}
\hat{\rho}_{in} = c_{vac}\ket{0}\bra{0} +  c_{sp}\ket{Q}\bra{Q} +  c_{tp}\ket{2}\bra{2},  
\label{E1}
\end{equation}
where $c_{vac}$ represents the vacuum component, $c_{sp}$ the single-photon component and $c_{tp}$ the two-photon component. We also model the phase noise, which will reduce our coherences to the expected average value of $0.276$. To do so, we take the qubit $\ket{Q}$ of eq. \ref{E1} ($\ket{Q}\bra{Q} = \ket{Q(\phi)}\bra{Q(\phi)}$) and model it as a mixture of states with a Gaussian phase distribution, centered around the desired qubit phase $\phi$,
\begin{equation}
\ket{Q(\phi)}\bra{Q(\phi)} ^\prime = \frac{1}{\sqrt{2 \pi}\Delta_{\phi}}\int_{-\pi}^{\pi}\ket{Q(\phi')}\bra{Q(\phi')}  e^{\frac{-(\phi-\phi')^2}{2\Delta_{\phi}^2}}d\phi', 
\end{equation}
where $\Delta_{\phi}$ is the standard deviation that we adapt to match the experimental value of the coherences. With this technique we can estimate the phase noise in our experiment to be around $11 \%$. With this phase noise estimation and the component values given in Table \ref{T2} we can thereby reproduce Table \ref{T1}. In this way the whole variety of experimentally realistic simulated input states is used to compute the fidelity with each experimental output state, in order to evaluate the minimal and maximal fidelity values we can expect. These values are given in table \ref{T3}. We then choose as experimental input state a state that gives us a fidelity the closest to the average and define the error bars as the distance to those absolute minimal and maximal values.

\section{Process Matrix Calculations}
Each quantum process can be described as a quantum map $\varepsilon(\hat{\rho})$ acting on the quantum system $\hat{\rho}$ \cite{1998TOMO}. In the process matrix representation the map is expressed via a $d^2\times d^2$ matrix $\chi$, with $d$ the dimension of the quantum system the map acts on, and decides on a full $d \times d$ basis $A_i$ such that the transformation is described by
\begin{equation}
\varepsilon(\hat{\rho}) = \sum_{n,m = 1}^{d^2}\chi_{nm}A_{m}\hat{\rho} A_n^\dagger
\label{equ1}
\end{equation}
where we expect $\chi$ to be a hermitian positive semidefinite matrix. If the process is trace preserving the condition $\sum_{n,m}\chi_{n,m}A_{m}^\dagger A_{n}= \mathbb{1}$ \cite{Baldwin2014} holds, marking the operation as a quantum channel \cite{weedbrook2012gaussian}. This condition holds for specific maps, for instance when the experimental process suffers from very low losses. In most cases, the trace is lowered homogeneously or inhomogeneously over the whole Bloch sphere, depending on the specific platform and basis. Our work corresponds to an inhomogeneous trace deformation as the teleportation becomes harder the more the input state approaches $\ket{1}$.

In order to reconstruct the matrix $\chi$ both the sets of states $\rho_{in}$ and measurements $E$ must form a basis of qubit density matrices, therefore each set holds $d^2$ elements. This gives the exact number of measurement on the exact number of states one has to perform in order to solve Eq. \ref{equ1} \cite{OtextquotesingleBrien2004}. In most cases we will want to measure a larger set, thereby over-defining the problem. Especially with state-dependent success probabilities and/or noise in the system, one cannot just select a data subset as the process reconstruction will differ heavily from sets to sets and therefore would change the reconstructed process matrix. To avoid this issue, the reconstruction is generally performed with the maximum-likelihood approach \cite{Fiurasek2001}. Unfortunately this approach is more difficult to implement for non-trace preserving processes like ours that do not have a constant scaling factor $\sum_{n,m}\chi_{n,m}A_{m}^\dagger A_{n}= c\mathbb{1}$. To avoid being trapped in local maxima
we choose to turn towards convex programming \cite{Baldwin2014} where we formulate a convex problem with no prior assumption on the process
\begin{align}
\text{minimize} \qquad &\sum_{jl}|f_{jl}-Tr(\sum_{n,m = 1}^{d^2}\chi_{nm}A_{m}\hat{\rho}_{in}^j A_n^\dagger E^l)|^2 \nonumber \\
\text{subject to} \qquad &\sum_{n,m}\chi_{n,m}A_{m}^\dagger A_{n}\leq \mathbb{1}  \nonumber \\
&\chi = \chi^\dagger \qquad \chi \geq 0 \nonumber
\end{align}
where one minimizes the square of the distance between the measured frequencies $f_{jl}=\textrm{Tr}(\hat{\rho}_{out}^j E^l)$ and the computed one given $\chi$. 
We use the language "Julia" with the package "Convex.jl" to translate the above formulated problem into a solvable convex form that is given to the solver "COSMO" \cite{Garstka2021}. 
With this method, we calculated our process matrix to be 
\begin{equation}
\chi = 
\begin{bmatrix}
 0.578+0.000i &    0.011-0.021i &   0.006-0.027i &    0.069-0.065i \\
 0.011+0.021i &    0.165-0.000i &   0.001-0.105i &    0.002-0.015i \\
 0.006+0.027i &    0.001+0.105i &   0.136-0.000i &   -0.010+0.006i \\
 0.069+0.065i &    0.002+0.015i &  -0.010-0.006i &    0.026-0.000i
\end{bmatrix}.
\end{equation}
where we chose the usual operator basis is the Pauli basis $\{\mathbb{1},\sigma_x,\sigma_y,\sigma_z\}$, connecting thereby the matrix elements to the fidelity of the process and combinations of bit and phase flip errors. We take the same sample size for each qubit. Using extrapolation methods we find that the maximal uncertainty on the process matrix elements is $\pm 0.0328$.

Comparing this process matrix to the ideal teleportation we find a process fidelity, meaning a fidelity between the ideal and experimental map, of $\mathcal{F}_{process}\simeq0.58$ - corresponding to $\chi_{11}$. This value can be connected to the average teleportation fidelity given pure input states $\bar{\mathcal{F}} = \frac{2\mathcal{F}_{process}+1}{3} = 0.72$ that would be obtained for a trace preserving processes \cite{Nielsen2002,Gilchrist2005}, thereby slightly overestimating our process. When applying the reconstructed process matrix to pure input states and then calculating the average teleportation fidelity we still beat the teleportation bound with $0.687>2/3$ albeit with a lower value then aforementioned, as expected in a non-trace-preserving process.

\section{Classical fidelity bound: allowing for mixed input qubits}
\label{NB}
The experimental input DV qubits are composed of a statistical mixture of a pure target qubit state $\ket{Q}$ with vacuum and multi-photon components, as written in Eq. \ref{E1}. Since the success of the protocol is assessed by the measurement of the fidelity between the input and the teleported qubits, it is important to consider this admixture, which can lead to an overestimation of the teleportation fidelity when the purity of the input qubit decreases. The standard teleportation fidelity bound is known as $\mathcal{F}=2/3$ \cite{OldF}, valid for pure states and when postselection is applied. It corresponds to the maximal fidelity in a teleportation setting without using an entangled pair. However, for instance if the vacuum component of the input qubit is always above $2/3$, one will always measure a fidelity above $2/3$ even if the teleportation does not occur and vacuum is created. 

The $\mathcal{F}=2/3$ bound is calculated assuming only pure input states, with therefore a Bloch vector radius of $|\vec{r}| = 1$. This bound is obtained by finding the optimal way of extracting information from a finite quantum ensemble with $N$ identical pure particles, which corresponds to the $2/3$ bound for $N=1$. No knowledge of the direction of those states on the Bloch sphere is assumed. Here we define a more general bound $\mathcal{F}^\prime$, used in the main text, that not only assumes zero knowledge of the direction of the states but also allows for mixed states with Bloch vector norm $|\vec{r}| \leq 1$. We take the approach presented in Ref. \cite{Bagan2006}, which is a direct generalization of Ref. \cite{OldF}.

We consider an input qubit density matrix written as
\begin{equation}
\hat{\rho} = \frac{1}{2}(\mathbb{1} + \vec{r}\,\cdot\,\vec{\sigma})
\end{equation}
where $\vec{r}=(r_x,r_y,r_z)$ is the Bloch vector and $\vec{\sigma}=(\boldsymbol{\sigma_x},\boldsymbol{\sigma_y},\boldsymbol{\sigma_z})$ a vector containing the Pauli matrices. We allow ourselves to act on this state with any positive operator valued measurement (POVM) $O_{\varepsilon}$ with outcomes $\varepsilon$ such that $ \sum_{\varepsilon}O_{\varepsilon} = \mathbb{1}$. This describes any possible local measurement as in Ref. \cite{OldF}. Based on the outcome, one will assume to have measured the state $\hat{\rho_{\varepsilon}}$ with the associated Bloch vector $\vec{R_{\varepsilon}}$. This assumption is measured via the fidelity between the actual state $\hat{\rho}$ and assumed state $\hat{\rho_{\varepsilon}}$
\begin{equation}
f(\vec{r},\vec{R_{\varepsilon}})  = \left(\textrm{Tr}(\sqrt{\sqrt{\hat{\rho_{\varepsilon}}}\hat{\rho}\sqrt{\hat{\rho_{\varepsilon}}}})\right)^2 .
\end{equation}

In order to take into account that the real state could have any possible direction and purity on the Bloch sphere we have to integrate over those degrees of freedom. Additionally all the possible operators $O_{\varepsilon}$ have to be taken into account such that the final quality of the assumption is computed as
\begin{equation}
F = \sum_{\varepsilon}\int d\rho f(\vec{r},\vec{R_{\varepsilon}}) p(\varepsilon|\vec{r}),
\label{Equ4}
\end{equation}
where the probability for the outcome $\varepsilon$ to occur is $p(\varepsilon|\vec{r})=\textrm{Tr}(O_{\varepsilon} \hat{\rho})$.

The prior probability distribution over the Bloch sphere, $d \rho$, has then to be defined. We assume to be completely ignorant regarding the direction of $\vec{r}$ and use the parametrisation $\vec{r} = (r\sin(\theta)\cos(\phi),r\sin(\theta)\sin(\phi),r\cos(\theta))$, with $|\vec{r}| = r$, $\theta \in [0,\pi]$ and $\phi \in [0,2\pi]$.
Our lack of knowledge of the direction implies $d \rho \propto d\Omega$ as the solid angle element $d\Omega$ represents an isotropical directional distribution on the Bloch sphere. We chose as radial prior the Bures prior $w(r) = \frac{4r^2}{\pi\sqrt{1-r^2}}$ \cite{Bagan2005}, as in Ref. \cite{Bagan2006,Osipov2010}, such that
\begin{equation}
d \rho = w(r) d\Omega d r = \frac{sin(\theta) r^2}{\pi^2\sqrt{1-r^2} } d\theta d\phi dr.
\end{equation}
In order to find the teleportation bound we have to maximize Eq. \ref{Equ4} for any POVM $O$. This is calculated by Ref. \cite{Bagan2006} to be possible when assuming a Bloch vector $\textbf{R}_{\varepsilon} := \textbf{V}_{\varepsilon} / |\textbf{V}_{\varepsilon}|$ with $\textbf{V}_{\varepsilon} = \int d\rho \textbf{r}p(\varepsilon|\vec{r})$ where  $\textbf{r}$ is the four-dimensional euclidean vector $\textbf{r} := (\sqrt{1-|\vec{r}|^2},\vec{r})$. Given this, we define the upper bound of Eq. \ref{Equ4} as the general teleportation bound 
\begin{equation}
F \xrightarrow{\text{maximize}}\mathcal{F} = 1/2(1+\sum_{\varepsilon}|\textbf{V}_{\varepsilon}|). \label{equ5} 
\end{equation}

With this general expression, we can first recover the result of Ref. \cite{OldF} for pure states by setting $|\vec{r}| = 1$, dropping thereby the integration over the Bloch radius for $\textbf{V}_{\varepsilon}$. In this case, the fidelity bound is indeed $\mathcal{F}^{|\vec{r}| = 1} = 2/3$. Equally we find the trivial result that $\mathcal{F}^{|\vec{r}| = 0} = 1$ for $|\vec{r}| = 0$ which shows that the parameters of fully mixed and therefore classical states can be extracted perfectly. These limiting cases give a good intuition about the teleportation bound: the more mixed quantum states are, the higher is the bound because it becomes easier to obtain the qubit parameters with a local measurement and thus sending information only via a classical channel gives larger fidelities.

Having gained this insight we can expect a mixed qubit bound in our case to be above the pure bound $\mathcal{F}^\prime > 2/3$. The produced input DV qubits in the $\{\ket{0}$, $\ket{1}\}$ basis will not exhibit all the same purity but a range of purities, depending on the weight of the single-photon component. In this case the minimal purity or Bloch sphere radius is defined via the escape efficiency of OPO IIa used to generate the qubits, i.e.,  $|\vec{r}|_{min} = |1-2\eta_{qubit}|=0.41$. We can therefore limit the integration space to $|\vec{r}| \in [|\vec{r}|_{min},|\vec{r}|_{max}]=[0.41,1]$ while renormalizing the Bures prior. This leads to the classical threshold used in this work given the input qubits, $\mathcal{F}^{|\vec{r}| \in [0.41,1]}= \mathcal{F}^\prime = 74.1\%$.

\section{Extension of our converter to dual-rail DV encodings }
\label{PIF}
Here we use the analysis of Takeda and coworkers \cite{Fuwa2013} about non-post-selected teleportation to preliminary assess the performance of our implementation with dual-rail discrete encodings such as time-bin or polarization. In non-post-selected schemes, vacuum and multiphoton components constitute errors that live in an orthogonal space to the qubit subspace, which has a weight $\eta$ given by the single-photon component of any created qubit. In this case, those error components can be viewed as classical and thus will be teleported with unit fidelity. The unconditional classical fidelity bound $\mathcal{F}_{\eta}$ calculated in \cite{Fuwa2013} is therefore the linear combination between detecting a classical state with probability $1-\eta$ and detecting the pure qubit with probability $\eta$, giving $\mathcal{F}_{\eta} = (1-\eta)*1+\eta*\frac{2}{3}=1-\frac{\eta}{3}$. This bound reaches back the $2/3$ value for pure states.

\begin{figure}[!h]
\centerline{\includegraphics{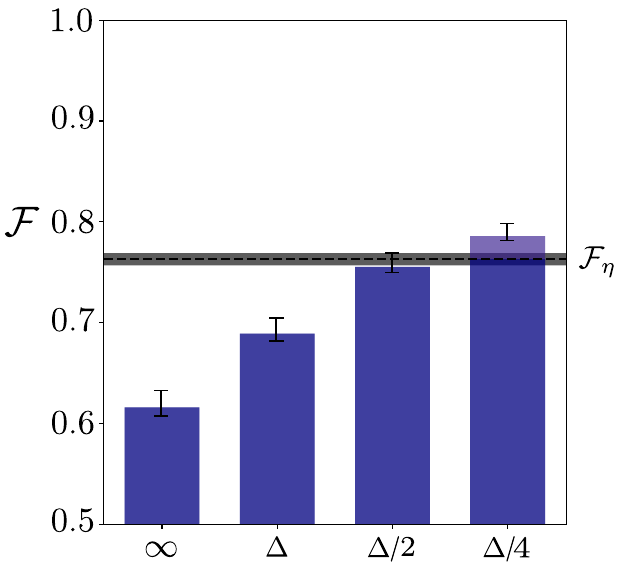}}
\vspace{-0.5cm}
\caption{\textbf{Fidelity for the conversion of the single-photon input state into an odd cat state as a function of the conditioning window of the BSM}. The shaded area represents the classical fidelity bound $\mathcal{F}_{\eta}$. $\infty$ denotes the absence of conditioning and $\Delta=\sigma_0$ is equal to the standard deviation of the vacuum shot noise.}
	\label{figBound}
\end{figure}

In our case the qubit fraction $\eta$ in the fidelity bound $\mathcal{F}_{\eta}$ would be given by the heralding efficiency of the single photon created by the straight-cut type-II OPO (OPO IIa). This heralding efficiency is measured independently and we obtain the value $\eta=71.2\%\pm1.5\%$, that yields $\mathcal{F}_{\eta}=76.3 \pm 0.5 \%$. As depicted in Fig. \ref{figBound} for the single-photon input, the only state in our work for which we can define the orthogonal error subspace, this bound is surpassed in our implementation with proper homodyne conditioning. The fidelity reaches $78.6^{+0.5}_{-0.1}\,\%>\mathcal{F}_{\eta}$ for a window equal to $\sigma_0/4$. This result constitutes a first assessment of our efficient implementation towards unconditional conversion for dual-rail encodings. For such implementation, the DV mode of the hybrid entangled state as well as the BSM should be adapted accordingly to the encodings. 

\end{document}